\documentclass[aps,prb,preprint,showpacs,superscriptaddress,floatfix]{revtex4}
\usepackage{graphics}
\usepackage{epsfig}
\usepackage{amsmath}
\usepackage{amssymb}
\usepackage{calc}
\newcommand{\beq}{\begin{equation}}
\newcommand{\eeq}{\end{equation}}
\newcommand{\bea}{\begin{eqnarray}}
\newcommand{\eea}{\end{eqnarray}}

\begin{document}
\title{
A three-qubit interpretation of BPS and non-BPS STU black holes}
\author{P\'eter L\'evay}
\affiliation{Department of Theoretical Physics, Institute of Physics, Budapest University of Technology and Economics, H-1521 Budapest, Hungary}
\date{\today}
\begin{abstract}
Following the recent trend we develop further the black hole analogy between quantum information theory and the theory of extremal stringy black hole solutions.
We show that the three-qubit interpretation of supersymmetric black hole solutions in the STU model can be extended also to include non-supersymmetric ones. First we show that the black hole potential can be expressed as one half the norm of a suitably chosen three-qubit entangled state containing the quantized charges and the moduli.
The extremization of the black hole potential in terms of this entangled state amounts to either supressing bit flip errors (BPS-case) or allowing very special types of flips transforming the states between different classes of non-BPS solutions.
We are illustrating our results for the example of the $D2-D6$ system.
In this case the bit flip errors are corresponding to sign flip errors of the   charges originating from the number of $D2$ branes. 
After moduli stabilization the states depending entirely on the charges are maximally entangled graph-states (of the triangle graph) well-known from quantum information theory.
An $N=8$ interpretation of the STU-model in terms of a mixed state with fermionic purifications is also given.

\end{abstract}
\maketitle{}

\section{Introduction}
In a recent series of papers\cite{Duff,Linde,Levay,Ferrara,Levay2,Duff2,Duff3} some interesting multiple relations have been established between quantum information theory and the physics of stringy black hole solutions. The activity in this field has started with the observation of Duff\cite{Duff} that the macroscopic black hole entropy for the BPS STU model can be expressed as an entanglement invariant characterizing three-qubit entanglement.
Later Kallosh and Linde\cite{Linde} have shown that for this model the different classes of black hole solutions correspond to the so called stochastic local operations and classical communication (SLOCC) classes of entanglement types characterizing three-qubit entanglement.
As a next step in our paper\cite{Levay} we have shown that the well-known process of moduli stabilization based on the supersymmetric attractor mechanism\cite{FKS,S,FK} in the entanglement picture corresponds to a distillation procedure of a GHZ-like state with maximum tripartite entanglement. 
This nice correspondence is based on the similar symmetry properties of the corresponding physical systems. For the STU model the symmetry group in question is $SL(2, {\bf R})^{\otimes 3}$ coming from the structure of the moduli space
$SL(2, {\bf R})/U(1)\times SL(2, {\bf R})/U(1)\times SL(2, {\bf R})/U(1)$, and for the quantum information theoretic scenario it is the group                   $SL(2, {\bf C})^{\otimes 3}$ related to the SLOCC group $GL(2, {\bf C})^{\otimes 3}$.  
Due to the very special structure of the STU model at first sight it seems that this black hole analogy should run out of steam for black hole solutions
corresponding to moduli spaces not exhibiting a product structure.
However, later work\cite{Ferrara,Levay2,Duff2} originating from the insight of Kallosh and Linde\cite{Linde}
revealed that the black hole entropy as a function of the $56$ charges ($28$ electric and $28$ magnetic) expressed in terms of Cartan's quartic invariant occurring in the context of $N=8$, $d=4$ supergravity\cite{Julia} with moduli space $E_{7(7)}/SU(8)$ can be understood as a special type of entangled system of seven qubits based on the discrete geometry of the Fano plane consisting of seven points and seven lines.
Moreover, different types of consistent truncations of this $N=8$, $d=4$ model in the entanglement picture can be understood via restriction to entangled subsets of the Fano plane. For example the STU model arising as a consistent truncation with $8$ charges (4 electric and 4 magnetic) is obtained by keeping merely one point of the Fano plane.
Based on recent results in the mathematics literature\cite{Elduque,Manivel} we conjectured\cite{Levay2} that  discrete geometric structures associated with the exceptional groups occurring in the magic square of Freudenthal and Tits might be related to interesting
entangled qubit systems which in turn can provide interesting connections to magic supergravities. In this context see also the interesting paper
 of Duff and Ferrara\cite{Duff3} streching the validity of the black hole analogy to the realm of black hole solutions in $d=5$ based on the group $E_{6(6)}$ connected to the bipartite entanglement of three qutrits.

Having turned out to be an ideal starting point for further generalizations in this paper our emphasis is once again on the STU model.
The reconsideration of this model is justified by the recent flurry of activity which has taken place with regards to the existence of non-BPS attractors\cite{Triv1,Triv2,Kallosh,Kallosh2}.
It turned out that as long as the black hole remains extremal many of the attractive features of the BPS solutions will survive in the non-BPS case too.
However, understanding the non-BPS case is considerably more difficult so in order to make progress as a first step it makes sense to restrict attention to a calculationally tractable subclass.
Motivated by these observations in this paper we would like to present a three-qubit interpretation of non-BPS
STU black holes in the hope that our observations might provide some additional insight into the development of this interesting field.

The plan of the paper is as follows. In Section II. background material
concerning extremal black hole solutions for ungauged $N=2$, $d=4$ supergravity coupled to some Abelian vector fields with special emphasis on the STU model is reviewed.
In Section III. we reintroduce a special type of three-qubit entangled state $\Psi$ first discussed in Ref. 3. This state is an unnormalized one containing $8$ complex amplitudes depending on the charges and the moduli. However as was discussed elsewhere\cite{Levay} , this state is $SU(2)^{\otimes 3}$ equivalent to a state with $8$ real amplitudes.  
The main result of this section is the elegant formula for the black hole potential expressed as one half the norm squared of this three-qubit state.
We show that the flat covariant derivatives with respect to the moduli are related to bit flip errors of $\Psi$. 
Armed with these results in Section IV. we reconsider the problem of extremization of the black hole potential in the context of quantum information theory.
We show that for BPS solutions at the attractor point  bit flip errors
on our three-qubit entangled state are supressed.
For non-BPS solutions the errors are not supressed but they are of very special form.
In order to gain some insight to the nature of these errors  we reconsider the solutions (D2-D6 system) studied by Kallosh et. al.\cite{Kallosh}
We show that for such systems the flips are transforming 
the states between different classes of solutions, and they are corresponding to sign flips of the charges $q_1,q_2$ and $q_3$ originating from the numbers of $D2$ branes.
After moduli stabilization the states depending entirely on the charges are 
maximally entangled graph-states (of the triangle graph) well-known from quantum    information
theory.
We also manage to embed the the $N=2$ STU model characterized by a three-qubit {\it pure} state to the $N=8$ one characterized by a three-qubit {\it mixed} one.
The conclusions and the comments are left for Section V.
In orther to make the paper self-contained, for the convenience of the reader some elementary results concerning graph states, bit flips, and Hadamard transforms are presented in the Appendix.

\section{STU black holes}

In the following we consider ungauged $N=2$ supergravity in $d=4$ coupled to
$n$ vector multiplets.
The $n=3$ case corresponds to the $STU$ model.
The bosonic part of the action (without hypermultiplets) is

\begin{eqnarray}
\label{Action}
{\cal S}&=&\frac{1}{16\pi}\int d^4x\sqrt{\vert g\vert }\{-\frac{R}{2}+G_{a\overline{b}}{\partial}_{\mu}z^a{\partial}_{\nu}{\overline{z}}^{\overline{b}}g^{\mu\nu}\nonumber\\&+&({\rm Im}{\cal N}_{IJ}{\cal F}^I{\cal F}^J+{\rm Re}{\cal N}_{IJ}{\cal F}^I{^\ast{\cal F}^J})\}
\end{eqnarray}
Here ${\cal F}^I$, and ${^\ast{\cal F}^I}$,  $I=1,2\dots n+1$ are two-forms associated to the field strengths ${\cal F}^I_{\mu\nu}$  of $n+1$ $U(1)$ gauge-fields and their duals.
The $z^a$ $a=1,\dots n$ are complex scalar (moduli) fields that can be regarded as local coordinates on a projective special K\"ahler manifold ${\cal M}$. This manifold for the STU model is
$SL(2, {\bf R})/U(1)\times SL(2, {\bf R})/U(1)\times SL(2, {\bf R})/U(1)$.
In the following we will denote the three complex scalar fields as

\beq
z^1\equiv S=S_1+iS_2,\qquad z^2\equiv T=T_1+iT_2,\qquad z^3\equiv U=U_1+iU_2.
\label{STU}
\eeq
\noindent
With these notations we have

\beq
G_{1\overline{1}}= G_{S\overline{S}}=\frac{1}{4S_2^2},\quad G_{2\overline{2}}=G_{T\overline{T}}=\frac{1}{4T_2^2},\quad G_{3\overline{3}}=G_{U\overline{U}}=\frac{1}{4U_2^2},
\label{targetmetric}
\eeq
\noindent
 with the other components like $G_{11}$, $G_{\overline{1}\overline{1}}$ and 
$G_{1\overline{2}}$ etc. are zero.
The metric above can be derived from the K\"ahler potential

\beq
K= -\log(-8U_2T_2S_2)
\label{Kahler}
\eeq
\noindent
as $G_{a\overline{b}}={\partial}_a{\partial}_{\overline{b}}K$.
In order to ensure the positivity of $e^K$ (needed later) we demand that
$S$, $T$ and $U$ should  have {\it negative} imaginary parts.

For the STU model the scalar dependent vector couplings ${\rm Re}{\cal N}_{IJ}$
and ${\rm Im}{\cal N}_{IJ}$ take the following form

\beq
{\rm Re}{\cal N}_{IJ}=\begin{pmatrix}2U_1T_1S_1&-U_1T_1&-U_1S_1&-T_1S_1\\-U_1T_1&0&U_1&T_1\\-U_1S_1&U_1&0&S_1\\-T_1S_1&T_1&S_1&0\end{pmatrix},
\label{valos}
\eeq 

\beq
{\rm Im}{\cal N}_{IJ}=U_2T_2S_2\begin{pmatrix}1+{\left(\frac{S_1}{S_2}\right)}^2+{\left(\frac{T_1}{T_2}\right)}^2+{\left(\frac{U_1}{U_2}\right)}^2&-\frac{S_1}{S_2^2}&-\frac{T_1}{T_2^2}&-\frac{U_1}{U_2^2}\\-\frac{S_1}{S_2^2}&\frac{1}{S_2^2}&0&0\\-\frac{T_1}{T_2^2}&0&\frac{1}{T_2^2}&0\\-\frac{U_1}{U_2^2}&0&0&\frac{1}{U_2^2}\end{pmatrix}.
\label{kepzetes}
\eeq
\noindent
We note that these scalar dependent vector couplings can be derived from the
holomorphic prepotential
\beq
F(X)=\frac{X^1X^2X^3}{X^0},\qquad X^I=(X^0,X^0z^a),
\eeq
\noindent
via the standard procedure characterizing special K\"ahler geometry\cite{Ceresole1}.
For the explicit expressions for ${\cal N}_{IJ}$ for general cubic holomorphic potentials see the recent paper of Ceresole et.al.\cite{Ceresole2}.

For the physical motivation of Eq.(\ref{Action}) we note that when type IIA string theory is compactified on a $T^6$ one recovers $N=8$ supergravity in $d=4$ with $28$ vectors and $70$ scalars taking values in the symmetric space $E_{7(7)}/SU(8)$. This $N=8$ model with an on shell U-duality symmetry $E_{7(7)}$ has a consistent $N=2$ truncation with $4$ vectors and three complex scalars which is just the STU model\cite{FK2}. 

Now we briefly recall the basic facts concerning static, spherically symmetric, extremal black hole solutions associated to the (\ref{Action}) action.
Let us consider the static spherically symmetric ansatz for the metric
\beq
ds^2=e^{2{\cal U}}dt^2-e^{-2{\cal U}}\left[\frac{c^4}{{\sinh}^4c\tau}d{\tau}^2+\frac{c^2}{{\sinh}^2c{\tau}}d{\Omega}^2\right],
\label{ansatz}
\eeq
\noindent
here ${\cal U}\equiv {\cal U}(\tau)$, $c^2=2{\cal S}{\cal T}$, where ${\cal S}$ is the entropy and ${\cal T}$ is the temperature of the black hole. The coordinate $\tau$ is a "radial" one, at infinity ($\tau\to 0$) we will be interested in solutions reproducing the Minkowski metric. $d{\Omega}^2$ is the usual metric of the unit two-spere in terms of polar coordinates $\theta$ and $\varphi$.
Our extremal black holes will correspond to the limit $c\to 0$ i.e. having vanishing Hawking temperature.
Putting this ansatz into (\ref{Action}) we obtain a one dimensional effective Lagrangian for the radial evolution of the quantities ${\cal U}(\tau)$, $z^a(\tau)$, as well as the electric ${\xi}^I(\tau)$, and magnetic ${\chi}_I(\tau)$ potentials defined as\cite{Kallosh2}

\beq
{\cal F}^I_{t\tau}={\partial}_{\tau}\xi^I,\qquad {\cal G}_{It\tau}\equiv
-i{\rm Im}{\cal N}_{IJ}\left(\ast{\cal F}\right)^J_{t\tau}-{\rm Re}{\cal N}_{IJ}{\cal F}^J_{t\tau}={\partial}_{\tau}{\chi}_I,
\eeq
\noindent

\beq
{\cal L}({\cal U}(\tau), z^a(\tau),\overline{z}^{\overline{a}}(\tau))=\left(    \frac{d{\cal U}}{d\tau}\right)^2+G_{a\overline{a}}\frac{dz^a}{d\tau}            \frac{d\overline{z}^{\overline{a}}}{d\tau}+e^{2{\cal U}}V_{BH}(z,\overline{z},p,q),
\eeq
\noindent
and the constraint

\beq
\left(
\frac{d{\cal U}}{d\tau}\right)^2+G_{a\overline{a}}\frac{dz^a}{d\tau}
\frac{d\overline{z}^{\overline{a}}}{d\tau}-e^{2{\cal U}}V_{BH}(z,\overline{z},p,
q)=c^2.
\label{constraint}
\eeq
Here our quantity of central importance is the Black Hole potential $V_{BH}$
is depending on the moduli as well on the quantized charges defined by

\beq
p^I=\frac{1}{4\pi}\int_{S^2}{\cal F}^I,\qquad
q_I=\frac{1}{4\pi}\int_{S^2}{\cal G}_I.
\eeq
\noindent
Its explicit form is given by

\beq
V_{BH}=\frac{1}{2}\begin{pmatrix}p^I&q_I\end{pmatrix}\begin{pmatrix}(\mu+\nu{\mu}^{-1}\nu)_{IJ}&-(\nu{\mu}^{-1})^J_I\\-({\mu}^{-1}\nu)^I_J&({\mu}^{-1})^{IJ}\end{pmatrix}\begin{pmatrix}p^J\\q_J\end{pmatrix},
\label{potential}
\eeq
\noindent
where the matrices ${\nu}={\rm Re}{\cal N}$ and ${\mu}={\rm Im}{\cal N}$ are the ones of Eqs. (5)-(6).
The explicit form of ${\mu}^{-1}$ is 

\beq
{\mu}^{-1}=\frac{1}{U_2T_2S_2}\begin{pmatrix}1&S_1&T_1&U_1\\S_1&{\vert S\vert}^2&S_1T_1&S_1U_1\\T_1&S_1T_1&{\vert T\vert}^2&T_1U_1\\U_1&S_1U_1&T_1U_1&{\vert U\vert}^2\end{pmatrix}.
\eeq
An alternative expression for $V_{BH}$ can be given in terms of the central charge of $N=2$ supergravity, i.e. the charge of the graviphoton (see e.g. Eq. (2.7) of Ref.16.).

\beq
V_{BH}=Z\overline{Z}+G^{a\overline{b}}(D_aZ)({\overline{D}}_{\overline{b}}\overline{Z})
\eeq
\noindent
where for the STU model

\beq
Z=e^{K/2}W=e^{K/2}(q_0+Sq_1+Tq_2+Uq_3+UTSp^0-UTp^1-USp^2-TSp^3),
\label{central}
\eeq
and $D_a$ is the K\"ahler covariant derivative 

\beq
D_aZ=({\partial}_a+\frac{1}{2}{\partial}_aK)Z.
\eeq
\noindent
Here $W(U,T,S)\equiv W(U,T,S;p,q)$ is the superpotential.

For extremal black hole solutions ($c=0$) the geometry is given by the line element

\beq
ds^2=e^{2{\cal U}}dt^2-e^{-2{\cal U}}\left[\frac{d{\tau}^2}{{\tau}^4}+\frac{1}{{\tau}^2}(d{\theta}^2+{\sin}^2\theta d{\varphi}^2)\right].
\eeq 
\noindent
The requirement for the solution to have finite horizon area is

\beq
e^{-2{\cal U}}\to \left(\frac{A}{4\pi}\right){\tau}^2,\qquad {\rm as}\qquad {\tau}\to -\infty,
\eeq  
\noindent
which using the new variable $r=-1/{\tau}$ is yielding for the near horizon geometry the $AdS_2\times S^2$ form

\beq
ds^2=\left(\frac{4\pi}{A}\right)r^2{dt^2}-\left(\frac{A}{4\pi}\right)\left[\frac{dr^2}{r^2}+(d{\theta}^2+{\sin}^2\theta d{\varphi}^2)\right].
\eeq

A particularly important subclass of solutions are the double-extremal solutions\cite{Wong,Behrndt,Lust}
These solutions have everywhere-constant moduli. These black holes pick up the
frozen values of the moduli that extremize the black hole mass  at infinity. The frozen values of the scalar fields are the ones at the horizon. These solutions are of Reissner-Nordstr\"om type with constant scalars defined by the critical point of the black
hole potential $V_{BH}$

\beq
{\partial}_aV_{BH}=0,\qquad z_{fix}(p,q)=z_{\infty}=z_{horizon}.
\eeq
\noindent
For such double-extremal black hole solutions the value of $A$ in Eq. (20) the area of the horizon is defined by the value of the black hole potential at the horizon\cite{Kallosh2}

\beq
\frac{A}{4\pi}=V_{BH}(z_{horizon},\overline{z}_{horizon},p,q).
\eeq
\noindent

Although our considerations in the following sections can obviously generalized for solutions of more general type\cite{Sabra1,Sabra2,Sabra3,Denef}, in order to simplify presentation in the following we restrict our attention to this particular subclass of double-extremal solutions.

\section{The Black Hole Potential as the norm of a three-qubit state}

In order to exhibit the interesting structure of the black hole potential Eq. (13) first we make some preliminary definitions.
As was observed by Duff\cite{Duff} it is useful to reorganize the charges of the STU model into the $8$ amplitudes of a three-qubit state

\beq
\vert\psi\rangle=\sum_{l,k,j=0}^1{\psi}^{lkj}\vert lkj\rangle\qquad \vert lkj\rangle\equiv\vert l\rangle_U\otimes\vert k\rangle_T\otimes\vert j\rangle_S
\eeq
\noindent
where
\beq
\begin{pmatrix}p^0,&p^1,&p^2,&p^3,&q_0,&q_1,&q_2,&q_3\end{pmatrix}=
\begin{pmatrix}{\psi}^{000},&{\psi}^{001},&{\psi}^{010},&{\psi}^{100},&-{\psi}^{111},&{\psi}^{110},&{\psi}^{101},&{\psi}^{011}\end{pmatrix} 
\eeq
\noindent
Notice , however that our identification of the amplitudes of the three-qubit state and the charges is slightly different from the one used by Duff\cite{Duff}.
Moreover, we have introduced the convention of labelling the qubits from the right to the left.   
Also we will regard the first, second and third qubits as the ones associated to some fictious subsystems $S$ (Sarah), $T$ (Tom), and $U$ (Ursula). 
The state $\vert\psi\rangle$ is a three-qubit state of a very special kind.
First of all this state defined by the charges need not have to be normalized. Moreover, the amplitudes of this state are not complex numbers but {\it integers}.
In the following we will refer to this state as the {\it reference state}. Now we are going to define a new unnormalized three-qubit state $\vert \Psi\rangle$ which is depending on the charges {\it and
also the moduli}\cite{Levay}. This new state will be a three-qubit state with $8$ complex amplitudes. However, as we will see it is really a {\it real three-qubit state},
since it is $SU(2)^{\otimes 3}$ equivalent to a one with $8$ real amplitudes\cite{Levay}. 

In order to motivate our definition of the new state $\vert\Psi\rangle$ we notice that

\beq
V_{BH}=-\frac{1}{2}\frac{1}{U_2T_2S_2}\langle\psi\vert\begin{pmatrix}{\vert U\vert}^2&-U_1\\-U_1&1\end{pmatrix}\otimes\begin{pmatrix}{\vert T\vert}^2&-T_1\\-T_1&1\end{pmatrix}\otimes\begin{pmatrix}{\vert S\vert}^2&-S_1\\-S_1&1\end{pmatrix}\vert\psi\rangle.
\eeq
\noindent
In order to prove this calculate the $8\times 8$ matrix in the middle with rows and columns labelled
in the binary form $000,001,010,011,100,101,110,111$, and regard $\vert\psi\rangle$ as the column vector $({\psi}^{000},{\psi}^{001},\dots,{\psi}^{111})^T$ and $\langle\psi\vert$ the corresponding row vector. It is straightforward to see that the resulting expression is the same as the one that can be obtained using Eqs. (5), (6), (13) and (14). For establishing this result note, however the different labelling
of rows and columns of matrices in Eqs. (13) (which is based on the symplectic structure ) and Eq. (25) (based on the binary labellig).  

Now we define the state $\vert\Psi\rangle$ as

\beq
\vert\Psi(U,\overline{U},T,\overline{T},S,\overline{S};p,q)\rangle =
e^{i\Phi}e^{K/2}\begin{pmatrix}\overline{U}&-1\\-U&1\end{pmatrix}\otimes
\begin{pmatrix}\overline{T}&-1\\-T&1\end{pmatrix}\otimes\begin{pmatrix}\overline{S}&-1\\-S&1\end{pmatrix}\vert\psi\rangle.
\label{Psi}
\eeq
\noindent
With the choice for the phase factor $e^{i\Phi}=e^{-3i\pi/4}$
the resulting matrices in the three-fold tensor pruduct are all $SL(2, {\bf C})$ ones. They are explicitly explicitly given by

\beq
{\cal A}_S\equiv\frac{e^{-i\pi/4}}{\sqrt{-2S_2}}\begin{pmatrix}\overline{S}&-1\\-S&1\end{pmatrix}=e^{i\pi/4}\frac{1}{\sqrt{2}}\begin{pmatrix}1&i\\1&-i\end{pmatrix}\frac{1}{\sqrt{-S_2}}\begin{pmatrix}-S_2&0\\-S_1&1\end{pmatrix}
\eeq

\beq
{\cal B}_T\equiv\frac{e^{-i\pi/4}}{\sqrt{-2T_2}}\begin{pmatrix}\overline{T}&-1\\-T&1\end{pmatrix}=e^{i\pi/4}\frac{1}{\sqrt{2}}\begin{pmatrix}1&i\\1&-i\end{pmatrix}\frac{1}{\sqrt{-T_2}}\begin{pmatrix}-T_2&0\\-T_1&1\end{pmatrix}
\eeq

\beq
{\cal C}_U\equiv\frac{e^{-i\pi/4}}{\sqrt{-2U_2}}\begin{pmatrix}\overline{U}&-1\\-U&1\end{pmatrix}=e^{i\pi/4}\frac{1}{\sqrt{2}}\begin{pmatrix}1&i\\1&-i\end{pmatrix}\frac{1}{\sqrt{-U_2}}\begin{pmatrix}-U_2&0\\-U_1&1\end{pmatrix}.
\eeq
\noindent
With this notation we have $\vert\Psi\rangle={\cal C}_U\otimes {\cal B}_T\otimes {\cal A}_S\vert\psi\rangle$.
This means that the states $\vert\Psi\rangle$ for all values of the moduli
are in the $SL(2, {\bf C})^{\otimes 3}$ orbit of the reference state $\vert\psi\rangle$  of Eq. (23) defined by the charges. This means that the value of the three-tangle ${\tau}_3$ (related to Cayley's hyperdeterminant) which is an $SL(2, {\bf C})^{\otimes 3}$ invariant
characterizing genuine three-qubit entanglement\cite{Kundu,Duff,Linde,Levay} is the same for both $\vert\psi\rangle$ and $\vert\Psi\rangle$.
Obviously the state $\vert\Psi\rangle$ is an unnormalized three-qubit one
with $8$ complex amplitudes. However, according to Eqs. (27-29) it is {\it not} a genuine complex three-qubit state but rather a one which is $SU(2)^{\otimes 3}$ equivalent to a real one. This should not come as a surprise 
since the symmetry group associated with the STU model is not $SL(2, {\bf C})^{\otimes 3}$ but rather $SL(2, {\bf R})^{\otimes 3}$.

Using Eq. (25) now we are ready to write the black hole potential in the following nice form

\beq
V_{BH}=\frac{1}{2}{\vert\vert\Psi\vert\vert}^2.
\eeq
\label{norma}
\noindent
Here the norm is defined using the usual scalar product in ${\bf C}^8\simeq {\bf C}^2\otimes {\bf C}^2\otimes{\bf C}^2$ with complex conjugation in the first factor. Since the norm is invariant under $U(2)^{\otimes 3}$ our choice of the phase factor $e^{i{\Phi}}$ is not relevant in the structure of $V_{BH}$.
In the following for the sake of calculational simplicity we set in Eq. (26) ${\Phi}\equiv 0$. However, in this convenient "gauge" the three-tangle ${\tau}_3$ for the charge-dependent $\vert\psi\rangle $ and the charge and moduli-dependent $\vert\Psi\rangle$ will no longer be the same.  
Hence the charge and moduli-dependent $\vert\Psi\rangle$ in the "gauge" $\Phi\equiv 0$ will be in the same SLOCC (i.e. $GL(2, {\bf C})^{\otimes 3}$ ) but not in the same $SL(2, {\bf C})^{\otimes 3}$ orbit as the charge-dependent reference state $\vert\psi\rangle$.
Moreover, we could have defined a new moduli dependent real state instead of the complex one $\vert\Psi\rangle$ by using merely the $SL(2, {\bf R})$ matrices
of Eqs. (27-29) for their definition.
However, we prefer the complex form of Eq. (26) since it will be useful later.

It is instructive to write out explicitly the amplitudes of our complex three-qubit state $\vert\Psi\rangle$.
After recalling the definition of the superpotential $W(U,T,S)$ of Eq. (16) they are

\beq
{\Psi}^{000}=e^{K/2}W(\overline{U},\overline{T},\overline{S}),\qquad {\Psi}^{111}=-e^{K/2}W(U,T,S),
\eeq

\beq
{\Psi}^{110}=e^{K/2}W({U},{T},\overline{S}),\qquad
 {\Psi}^{001}=-e^{K/2}W(\overline{U},\overline{T},S),
\eeq

\beq
{\Psi}^{101}=e^{K/2}W({U},\overline{T},{S}),\qquad
 {\Psi}^{010}=-e^{K/2}W(\overline{U},T,\overline{S}),
\eeq

\beq
{\Psi}^{011}=e^{K/2}W(\overline{U},{T},{S}),\qquad
 {\Psi}^{100}=-e^{K/2}W(U,\overline{T},\overline{S}).
\eeq
\noindent
We can summarize this as

\beq
{\Psi}^{lkj}=(-1)^{l+k+j}e^{K/2}W^{lkj}, \quad {\rm where}\quad W^{101}\equiv W(U,\overline{T},S)\quad{\rm etc.}.
\eeq
Notice also that we have the property

\beq
{\Psi}^{000}=-\overline{{\Psi}^{111}},\quad 
{\Psi}^{110}=-\overline{{\Psi}^{001}},\quad
{\Psi}^{101}=-\overline{{\Psi}^{010}},\quad
{\Psi}^{011}=-\overline{{\Psi}^{100}}.
\eeq
\noindent 
Using this in Eq. (30) we can write $V_{BH}$ in the alternative form

\beq
V_{BH}=e^K\left({\vert W(U,T,S)\vert}^2+{\vert W(U,T,\overline{S})\vert}^2+
{\vert W(U,\overline{T},S)\vert}^2+{\vert W(\overline{U},T,S)\vert}^2\right),
\eeq
\noindent
in agreement with the result found in Eq. (A.39) of the Appendix of Ref. 16.

As a next step we would like to clarify the meaning of the complex amplitudes ${\Psi}^{lkj}$. For this we have to look at the structure of covariant derivatives.
Using Eq. (17) we have
$D_aW={\partial}_aW+({\partial}_aK)W$
so for example

\beq
D_SW(U,T,S)=\frac{W(U,T,\overline{S})}{\overline{S}-S} 
\eeq
\noindent
Since the nonzero components of the Christoffel symbols are

\beq
{\Gamma}^S_{SS}=\frac{2}{\overline{S}-S},\quad
{\Gamma}^T_{TT}=\frac{2}{\overline{T}-T},\quad                                  {\Gamma}^U_{UU}=\frac{2}{\overline{U}-U},
\eeq
\noindent
we have\cite{Kallosh}

\beq
D_SD_TW(U,T,S)=\frac{W(U,\overline{T},\overline{S})}{(\overline{S}-S)(\overline{T}-T)},\quad
D_SD_SW(U,T,S)=0\quad {\rm etc.}
\eeq
\noindent

It is convenient to introduce flat covariant derivatives.
Let ${\delta}_{\hat{a}\hat{\overline{b}}}$ be the flat Euclidean metric.
Then we define the vielbein $e_a^{\hat{a}}$ via the expression $G_{a\overline{b}}=e^{\hat{a}}_ae^{\hat{\overline{b}}}_{\overline{b}}{\delta}_{\hat{a}\hat{\overline{b}}}$.
Using Eq. (3) we get for the nonzero components of the {\it inverse} of $e_{\hat{a}}^a$

\beq
e_{\hat{S}}^S=i(S-\overline{S})=-2S_2,\quad
e_{\hat{T}}^T=i(T-\overline{T})=-2T_2,\quad
e_{\hat{U}}^U=i(U-\overline{U})=-2U_2.
\eeq
\noindent
The flat covariant derivatives are defined by
$D_{\hat{a}}=e_{\hat{a}}^aD_a$.
Using Eqs. (31-34) and (38) we see that

\beq 
D_{\hat{S}}{\Psi}^{111}=i{\Psi}^{110},\quad
D_{\hat{T}}{\Psi}^{111}=i{\Psi}^{101},\quad
D_{\hat{U}}{\Psi}^{111}=i{\Psi}^{011},
\eeq

\beq
D_{\hat{\overline{S}}}{\Psi}^{111}=D_{\hat{\overline{T}}}{\Psi}^{111}=
D_{\hat{\overline{U}}}{\Psi}^{111}=0.
\eeq
\noindent
It is straightforward to verify that the action of the operators $D_{\hat{a}}$
and $D_{\hat{\overline{a}}}$ on the remaining amplitudes follows the same pattern.
We can neatly summarize their action after defining the raising and lowering operators $S_{\pm}$

\beq
S_{+}\vert 0\rangle=\vert1\rangle,\qquad S_{+}\vert 1\rangle =0,\qquad S_{-}\vert 0\rangle=0,\qquad S_{-}\vert 1\rangle=\vert 0\rangle.
\eeq 
\noindent
Hence the flat covariant derivatives are transforming between the $8$ amplitudes
${\Psi}^{lkj}$
and the combinations like $I\otimes I\otimes S_{\pm}$ are transforming between the $8$ basis vectors $\vert lkj\rangle$ of the three qubit state $\vert\Psi\rangle$.
In fact one can verify that

\beq
\frac{1}{i}D_{\hat{S}}\vert\Psi\rangle=(I\otimes I\otimes S_{+})\vert\Psi\rangle,\qquad -\frac{1}{i}D_{\hat{\overline{S}}}\vert\Psi\rangle=
(I\otimes I\otimes S_{-})\vert\Psi\rangle,
\eeq

\beq
\frac{1}{i}D_{\hat{T}}\vert\Psi\rangle=(I\otimes S_{+}\otimes I)\vert\Psi\rangle
,\qquad -\frac{1}{i}D_{\hat{\overline{T}}}\vert\Psi\rangle=
(I\otimes  S_{-}\otimes I)\vert\Psi\rangle,
\eeq

\beq
\frac{1}{i}D_{\hat{U}}\vert\Psi\rangle=(S_{+}\otimes I\otimes I )\vert\Psi      \rangle
,\qquad -\frac{1}{i}D_{\hat{\overline{U}}}\vert\Psi\rangle=
(S_{-}\otimes I\otimes I)\vert\Psi\rangle.
\eeq
\noindent
Hence the flat covariant derivatives are acting on our three-qubit state $\vert\Psi\rangle$ as the operators of {\it projective errors} known from the theory of quantum error correction (see Appendix).
Alternatively one can look at the action of the combination
$(D_{\hat{a}}-D_{\hat{\overline{a}}})/i$

\beq
\frac{1}{i}(D_{\hat{S}}-D_{\hat{\overline{S}}})\vert\Psi\rangle =(I\otimes I\otimes X)\vert\Psi\rangle\quad{\rm etc.}
\eeq
\noindent
where $I\otimes I\otimes X$ is the operator of bit-flip error acting on the {\it first qubit} (see Appendix).

Having clarified the meaning of the entangled three-qubit state $\vert\Psi\rangle$ and the flat covariant derivatives as error operations acting on it, in light of these result  in the next section we would like to obtain some additional insight on the structure of BPS and non-BPS black hole solutions.

\section{BPS and non-BPS solutions}

As it is well-known\cite{Kallosh,Kallosh2,Vafa}
the extremization of the black-hole potential Eq. (15) with respect              to the moduli
yields the following set of equations

\beq
{\partial}_{a}V_{BH}=e^K\left(G^{b\overline{c}}(D_aD_bW)\overline{D}_{\overline{c}}\overline{W}+2(D_aW)\overline{W}\right)=0,
\eeq

\beq
{\partial}_{\overline{a}}V_{BH}=e^K\left(G^{\overline{b}c}(D_{\overline{a}}D_{\overline{b}}\overline{W})D_cW+2(\overline{D}_{\overline{a}}\overline{W})W\right)=0.
\eeq
\noindent
Assuming $W\neq 0$ expressing $\overline{D}_{\overline{a}}\overline{W}$ 
from Eq. (50), and substituting the resulting expression back to Eq. (49)
yields an equation\cite{Vafa} of the form

\beq
M_a^b(D_bW)=0.
\eeq
\noindent
For the STU-model using Eq. (3), and Eq. (42) with its complex conjugate
for the matrix $M$ we get the following expression 

\beq
M_a^b=\begin{pmatrix}4{\Psi}^7{\Psi}^0-{\Psi}^4{\Psi}^3-{\Psi}^5{\Psi}^2&\frac{T_2}{S_2}{\Psi}^6{\Psi}^2&\frac{U_2}{S_2}{\Psi}^6{\Psi}^4\\\frac{S_2}{T_2}{\Psi}^5{\Psi}^1&4{\Psi}^7{\Psi}^0-{\Psi}^6{\Psi}^1-{\Psi}^4{\Psi}^3&\frac{U_2}{T_2}{\Psi}^5{\Psi}^4\\ \frac{S_2}{U_2}{\Psi}^3{\Psi}^1&\frac{T_2}{U_2}{\Psi}^3{\Psi}^2&4{\Psi}^7{\Psi}^0-{\Psi}^6{\Psi}^1-{\Psi}^5{\Psi}^2\end{pmatrix},
\eeq
\noindent
where for simplicity we used the decimal notation $({\Psi}^{000},\dots,{\Psi}^{111})=({\Psi}^0,,\dots{\Psi}^7)$.
Expressing the covariant derivatives $D_aW$ in terms of the corresponding amplitudes using Eq. (42) (related to the flat covariant ones), we obtain the explicit expression for Eq. (51)

\beq
(2{\Psi}^7{\Psi}^0-{\Psi}^5{\Psi}^2-{\Psi}^4{\Psi}^3){\Psi}^6=0
\eeq
\beq
(2{\Psi}^7{\Psi}^0-{\Psi}^4{\Psi}^3-{\Psi}^6{\Psi}^1){\Psi}^5=0
\eeq
\beq
(2{\Psi}^7{\Psi}^0-{\Psi}^6{\Psi}^1-{\Psi}^5{\Psi}^2){\Psi}^3=0.
\eeq
\noindent
Recall also that ${\Psi}^{7-\alpha}=-\overline{\Psi}^{\alpha}$ where $\alpha=0,1\dots 7$ which is just the decimal form of Eq. (36).
The determinant of $M_a^b$ is

\beq
\frac{1}{4}{\rm Det}M=\vert{\Psi}^0\vert^2(4\vert{\Psi}^0\vert^2-\vert{\Psi}^1\vert^2-\vert{\Psi}^2\vert^2-\vert{\Psi}^4\vert^2)-\vert{\Psi}^1{\Psi}^2{\Psi}^4\vert^2.
\eeq
\noindent
Using these results we can conclude that there are two different types of solutions for $Z\neq 0$.

{\bf I. BPS solutions}

\beq
{\Psi}^1={\Psi}^2={\Psi}^4=0, \qquad {\rm Det}M\neq0.
\eeq

\noindent

{\bf II. Non-BPS solutions} 

\beq
\vert{\Psi}^0\vert^2=\vert{\Psi}^1\vert^2=\vert{\Psi}^2\vert^2=\vert{\Psi}^4\vert^2,\qquad {\rm Det}M=0.
\eeq

Notice that the amplitudes  ${\Psi}^0={\Psi}^{000}$ and ${\Psi}^7={\Psi}^{111}$ are playing a special role in the STU model. Indeed they are related to the cenral charge and its complex conjugate as

\beq
Z=-{\Psi}^{7},\qquad \overline{Z}={\Psi}^{0}.
\eeq
\noindent
For the type of solutions considered here $Z\neq 0$, hence the corresponding amplitudes are never zero. We should remark, however at this point that there are
solutions belonging to a third class\cite{Bellucci1,Bellucci2}: the ones with $Z\equiv 0$.
The structure of these solutions 
has recently been studied in the context of the STU-model\cite{Bellucci3}.
In the next sections we are focusing merely on classes I. and II. where an interpretation of known results in the language of quantum information theory is 
straightforward. It is easy to extend our considerations also to the third class however, we postpone the investigation of these solutions for the special case of the $D2-D6$ system untill Section V.  
Untill then let us try to find a quantum information theoretic interpretation for the two types of solutions found above.

\subsection{BPS solutions}
We know that for BPS black holes at the horizon ($r=0$) we have $D_aZ\equiv 0$.
From Eq. (42-43) and their complex conjugates we see that
the only non-vanishing amplitudes of $\vert\Psi\rangle$ at the horizon are 
${\Psi}^{000}$ and ${\Psi}^{111}$, hence for the BPS case

\beq
\vert\Psi(0)\rangle=\overline{Z}\vert 000\rangle -Z\vert 111\rangle.
\eeq
\noindent
This state is of the generalized GHZ (Greenberger-Horne-Zeilinger) form of maximal tripartite entanglement (See Ref. 3 and references therein).
The form of the black hole potential at the horizon  is
\beq
V_{BH}=\frac{1}{2}(\vert{\Psi}^{000}\vert^2+\vert{\Psi}^{111}\vert^2)=\vert Z\vert^2=M^2_{BPS}.
\eeq
Notice, that for double-extremal black holes Eq. (60-61) are valid even away from the horizon. However, for BPS solutions of more general type $\vert\Psi\rangle$ as a function of $\tau$ ( or $r$) is of the general form of Eq. (26)
with the moduli $S(\tau)$ , $T(\tau)$ and $U(\tau)$ being solutions for the equations of motion for the moduli\cite{Sabra1,Sabra2,Sabra3,Denef}.
Of course these solutions at the horizon ($r=0$, $\tau=-\infty$) will again be attracted to the very special form of $\vert\Psi\rangle$ as dictated by Eqs. (60-61).
Hence the first interpretation of the attractor mechanism for the BPS case
is that of a quantum information theoretic distillation of a GHZ-like state
Eq. (60) at the horizon from a one of the general form Eq. (26).
As we reach the horizon
the conditions
\beq
D_{S}Z=D_{T}Z=D_{U}Z=0,\qquad Z\neq 0.
\eeq
\noindent 
guarantee that
\beq
{\Psi}^{110}={\Psi}^{101}={\Psi}^{011}={\Psi}^{001}={\Psi}^{010}={\Psi}^{100}=0,
\eeq
\noindent
hence we are left merely with the GHZ components ${\Psi}^{000}$ and ${\Psi}^{111}$.

Eqs. (63) can be used to express the stabilized values of the moduli in terms of the charges\cite{Behrndt}. For a geometric and quantum information theoretic reinterpretation and the explicit form of these solutions see Ref 3.
Here we merely cite the result for the form of the three-qubit entangled state at the horizon ($\varrho=0$) 

\beq
\label{verynew}
\vert{\Psi}(0)\rangle =(- D)^{1/4}[e^{i\delta}\vert 000\rangle -e^{-i\delta}\vert 111\rangle].
\eeq
\noindent
Here
\begin{widetext}
\beq
D=(p\circ q)-4((p^1q_1)(p^2q_2)+(p^1q_1)(p^3q_3)+(p^2q_2)(p^3q_3))+4p^0q_1q_2q_3-4q_0p^1p^2p^3,
\eeq
\end{widetext}
\noindent
where
$p\circ q=p^0q_0+p^1q_1+p^2q_2+p^3q_3$,  is Cayley's hyperdeterminant\cite{Duff, Cayley},
and

\beq
\tan\delta =\sqrt{-D}\frac{p^0}{2p^1p^2p^3-p^0(p\circ q)}.
\eeq
For the BPS solution to be consistent we have to require $-D>0$ otherwise the scalar fields are real and the K\"ahler potential is not defined.
Using Eqs. (22) and (61) and the Bekenstein-Hawking entropy formula ${\cal S}=A/4$ 
we get the well-known result

\beq
{\cal S}=\pi\sqrt{-D}.
\eeq
\noindent
Notice, however that apart from reproducing the result of Ref. 22 we have also calculated a useful quantity namely our entangled three-qubit state at the horizon. As we will see in the following this quantity will give us extra information on the nature of both BPS and non-BPS solutions.

As an important special case (to be also discussed later in the non-BPS context) let us consider the $D2-D6$ system\cite{Kallosh}. In this case $q_0=p^1=p^2=p^3=0$ and the superpotential is of the form $W=UTSp^0+Sq_1+Tq_2+Uq_3$. Using Eqs. (64-66) the three-qubit entangled state at the horizon is 

\beq
\vert\Psi(0)\rangle=i\sqrt{2}(-p^0q_1q_2q_3)^{1/4}(\vert 000\rangle +\vert 111\rangle),
\eeq
\noindent
where $-p^0q_1q_2q_3>0$. Eq. (68) is just the (unnormalized) canonical GHZ-state. Notice that for the charge dependent reference state $\vert\psi\rangle$ of Eq. (23)
$D(\psi)=4p^0q_1q_2q_3<0$, but $D(\Psi(0))=({\Psi}^{000}{\Psi}^{111})^2=-4p^0q_1q_2q_3>0$. This change of sign is due to our choice of "gauge" ${\Phi}\equiv 0$ in Eq. (26). 
(See the discussion after Eq. (30).)
From Eq. (60) we see that for this  $D2-D6$ system the value of the central charge at the horizon is in accordance with Eq. (4.14) of Ref. 19 is $Z=-i\sqrt{2}(-p^0q_1q_2q_3)^{1/4}$.

Untill this point we have discussed a quantum information theoretic reinterpretation of the attractor mechanism for BPS black-hole solutions.
In this picture we are looking at the dynamical system as a one starting from the asymptotically Minkowski geometry where $\vert\Psi(r)\rangle$ is of the general form (Eq. (26)), and when reaching the horizon with $AdS_2\times S^2$ geometry one is left with $\vert\Psi(0)\rangle$  a GHZ-like state.

However we have an alternative way of interpretation.
In this picture one is starting from the {\it horizon} with the state $\vert\Psi(0)\rangle$.
We know that this state is of the GHZ (i.e. maximally entangled) form of Eq. (60).
According to Eqs. (62) and the interpretation of the action of the flat covariant derivatives as error operators (see Eq. (48)) we see that in the BPS case
our GHZ-state $\vert\Psi(0)\rangle$ is protected from bit flip errors.
The BPS conditions of Eq. (62) are precisely the ones of supressing the bit flip errors for the three-qubit state $\vert\Psi(0)\rangle$  characterizing the extremal BPS black-hole solution.
Notice also that bit flips in the computational base correspond to phase flips in the Hadamard transformed base (see Appendix). By writing Eq. (68)
in the Hadamard transformed base we get

\beq
\vert\Psi(0)\rangle=i(-p^0q_1q_2q_3)^{1/4}[\vert\overline{0}\overline{0}\overline{0}\rangle+\vert\overline{0}\overline{1}\overline{1}\rangle+\vert\overline{1}\overline{0}\overline{1}\rangle+\vert\overline{1}\overline{1}\overline{0}\rangle].
\eeq
\noindent
Hence the observation that for the state $\vert\Psi(0)\rangle$ bit flip errors in the computational base are supressed  also means that 
errors of the form  
\beq
(I\otimes I\otimes X)\vert\Psi(0)\rangle=i(-p^0q_1q_2q_3)^{1/4}[\overline{0}\overline{0}\overline{0}\rangle-\vert\overline{0}\overline{1}\overline{1}\rangle-\vert\overline{1}\overline{0}\overline{1}\rangle+\vert\overline{1}\overline{1}\overline{0}\rangle]
\eeq
\noindent
changing the relative phase of the states in the Hadamard transformed base are not allowed.
Moreover it is instructive to consider the state Eq. (69) together with the "reference" state $\vert
\psi\rangle$ which is also depending merely on the charges

\beq
\vert\psi\rangle=p^0\vert 000\rangle+q_3\vert 011\rangle +q_2\vert 101\rangle   +
q_1\vert 110\rangle.
\eeq
\noindent
Hence for the $D2-D6$ system the charge dependent state resulting from moduli stabilization (Eq. (69)) is arising from the reference state (Eq. (71)) via discrete Fourier (Hadamard) transformation and uniformization of the amplitudes.
Moreover a comparison of Eq. (70) with Eq. (71) suggests that these bit flip errors are somehow connected to sign flip errors of the charges corresponding to $D2$ branes. This conjecture will be verified in the next subsection.  

\subsection{Non-BPS solutions.}
In order to gain some insight into the structure of non-BPS solutions
provided by quantum information theory we consider the specific example of the $D2-D6$ system also studied in the previous subsection. This system was studied in detail in Ref. 16.
By minimizing the effective potential the solutions to the moduli are\cite{Kallosh}

\beq
S=\pm i\sqrt{\frac{q_2q_3}{p^0q_1}},\qquad
T=\pm i\sqrt{\frac{q_1q_3}{p^0q_2}},\qquad
U=\pm i\sqrt{\frac{q_1q_2}{p^0q_3}},\qquad p^0q_1q_2q_3>0,
\eeq
\noindent
where the sign combinations not violating the positivity of $e^K$ are

\beq 
\{(-,-,-), (-,+,+), (+,-,+), (+,+,-)\}.
\eeq
\noindent
In Ref. 16. it was also checked that  these solutions are also forming  stable attractors, meaning that the extremum of the black hole potential is also a minimum.
In the following we would like to use these solutions to calculate $\vert\Psi(0)\rangle$ and study its behavior with respect to bit flip errors.

For the $(-,-,-)$ class straightforward calculation gives the result

\beq
{\Psi}^{000}={\Psi}^{111}=-\frac{i}{\sqrt{8}}(p^0q_1q_2q_3)^{1/4}\{ {\rm sgn}(p^0)-{\rm sgn}(q_3)-{\rm sgn}(q_2)-{\rm sgn}(q_1)\} 
\eeq

\beq
{\Psi}^{011}={\Psi}^{100}=-\frac{i}{\sqrt{8}}(p^0q_1q_2q_3)^{1/4}\{ {\rm sgn}(p^
0)-{\rm sgn}(q_3)+{\rm sgn}(q_2)+{\rm sgn}(q_1)\}
\eeq

\beq
{\Psi}^{101}={\Psi}^{010}=-\frac{i}{\sqrt{8}}(p^0q_1q_2q_3)^{1/4}\{ {\rm sgn}(p^
0)+{\rm sgn}(q_3)-{\rm sgn}(q_2)+{\rm sgn}(q_1)\}
\eeq

\beq
{\Psi}^{110}={\Psi}^{001}=-\frac{i}{\sqrt{8}}(p^0q_1q_2q_3)^{1/4}\{ {\rm sgn}(p^
0)+{\rm sgn}(q_3)+{\rm sgn}(q_2)-{\rm sgn}(q_1)\}.
\eeq
\noindent
For definitness we consider the case $p^0>0, q_1>0, q_2>0, q_3>0$ which is compatible with the constraint $p^0q_1q_2q_3>0$.
In this case we obtain the state $\vert\Psi(0)\rangle$

\beq
{\vert\Psi(0)\rangle}_{---}=\omega\{\vert 000\rangle -\vert 001\rangle -\vert 010\rangle -\vert 011\rangle -\vert 100\rangle -\vert 101\rangle -\vert 110\rangle+\vert 111\rangle\}.
\eeq
\noindent
where
\beq
\omega =\frac{i}{\sqrt{2}}(p^0q_1q_2q_3)^{1/4}.
\eeq
\noindent
From this state we see that $Z=-{\Psi}^{111}=-\omega$ in agreement with Eq. (4.16) of Ref. 19. 
In the Hadamard transformed basis this state takes the form
\beq
{\vert\Psi(0)\rangle}_{---}=-i(p^0q_1q_2q_3)^{1/4}\{\vert\overline{0}\overline{0}\overline{0}\rangle -\vert\overline{011}\rangle -\vert\overline{1}\overline{0}\overline{1}\rangle -\vert \overline{110}\rangle\}.
\eeq
\noindent
Comparing Eq. (69) and (80) we see that the basic difference between the BPS and non-BPS case is the change of sign in the combination $p^0q_1q_2q_3$ and also
 the appearance of a nontrivial relative phase between the Hadamard transformed basis vectors.

Let us now consider the class $(-,+,+)$.
Since for the $(-,-,-)$ class we had $S_2=-{\rm sgn}(q_1)\sqrt{q_1q_2q_3/p^0}$,
$T_2=-{\rm sgn}(q_2)\sqrt{q_1q_2q_3/p^0}$ and ,$U_2=-{\rm sgn}(q_3)\sqrt{q_1q_2q_3/p^0}$ then going from the class $(-,-,-)$ to the one of $(-,+,+)$ 
amounts to changing the signs of $q_1$ and $q_2$. (Remember our convention of labelling everything from the right to the left.)
As a result according to Eq. (75) the amplitudes ${\Psi}^{011}$ and ${\Psi}^{100}$
will be positive and the remaining ones are negative. The resulting state in this case is of the form

\beq
\vert{\Psi}(0)\rangle=\omega\{-\vert 000\rangle -\vert 001\rangle -\vert 010\rangle +\vert 011\rangle+\vert 100\rangle -\vert 101\rangle -\vert 110\rangle -\vert 111\rangle\}
\eeq
\noindent
or in the Hadamard transformed base
\beq
{\vert{\Psi}(0)\rangle}_{-++}=-i(p^0q_1q_2q_3)^{1/4}\{\vert\overline{000}\rangle -\vert \overline{011}\rangle +\vert\overline{101}\rangle +\vert\overline{110}\rangle\}.
\eeq
\noindent
We can summarize these observations for all classes of non-BPS attractors with $Z\neq 0$ for the $D2-D6$ system as

\beq
{\vert{\Psi}(0)\rangle}_{\gamma\beta\alpha}=-i(p^0q_1q_2q_3)^{1/4}\{\vert\overline{000}\rangle+\gamma\vert\overline{011}\rangle +\beta\vert\overline{101}\rangle+\alpha\vert\overline{110}\rangle\},
\eeq
\noindent
where $(\gamma,\beta,\alpha)=\{(-,-,-),(-,+,+),(+,-,+),(+,+,-)\}$.
Notice also that for example

\beq
(I\otimes I\otimes X){\vert{\Psi}(0)\rangle}_{---}={\vert{\Psi}(0)\rangle}_{++-},\qquad
(I\otimes I\otimes X){\vert{\Psi}(0)\rangle}_{-++}={\vert{\Psi}(0)\rangle}_{+-+},\qquad{\rm e.t.c}.
\eeq
\noindent
This means that the bit flip operators $I\otimes I\otimes X$, $I\otimes X\otimes I$ and $X\otimes I\otimes I$ are transforming in between the admissible classes of Eq. (73). The rule of transformation is: those class labels that are in the same slot as the bit flip operator $X$ {\it are not changed}, while the remaining ones are flipped.

What about physics?
The non-BPS black holes corresponding to attractors of a $D2-D6$ system
can be characterized by the four three-qubit entangled states of Eq. (83)
depending merely on the charges.
This equation should be taken together with the other charge-dependent state
of Eq. (71). It is clear from Eq. (84) that the error operation on the {\it first} qubit which is changing the signs of $\beta$ and $\gamma$ (the entries of the {\it second and third} slots) 
is corresponding to  a sign change of $q_2$ and $q_3$ (the {\it second} and {\it third}) of the charges in the reference state Eq. (71). 
Generally a bit flip error on the $j$th qubit corresponds to a sign flip of the $k$th an $l$th charge $q_k$ and $q_l$ where $j\neq k\neq l$, and $j,k,l=1,2,3$.

At this point we can obtain an additional insight into the BPS case as well.
Looking back at Eq. (70) which again corresponds to sign flips of charges $q_2$ and $q_3$, we understand that in the BPS case {\it sign flips of these kind are supressed}. Although these sign flips are not changing the sign of the combination $p^0q_1q_2q_3$ they are not allowed due to supersymmetry.
On the other hand
for non-BPS black holes flipping the signs of a pair of charges corresponding to changing the sign of the number of $D2$ branes. (Negative number of branes correspond to positive number of antibranes of the same kind.)
These transformations can be regarded as bit flip errors transforming one non-BPS solution to the other. 
Moreover according to Eqs. (45-48) we also see that these bit flip errors have their origin in the action of the flat covariant derivatives on our moduli dependent entangled state of Eq. (26).

In closing this section we make an additional interesting observation.
As we have already realized for the BPS case, at the horizon the form of the
three-qubit entangled state will be of very special form. For the $D2-D6$ system it is proportional to the canonical $GHZ$ state. 
What about the non-BPS case? Comparing Eqs. (68) for the BPS and (78) for the non-BPS $(-,-,-)$-class we see that unlike the GHZ state Eq. (78) does not seem to be related to any three-qubit state of special importance in quantum information theory. However, the state of Eq. (78) is a particularly nice example of a {\it graph-state}\cite{Dur}. Graph states are under intense scrutiny these days
due to the special role they are playing in quantum error correction, and in the study of correlations in wave functions of many body systems. 
Here we would like to show that the non-BPS states associated to the classes
of Eq. (73) are (unnormalized) graph states based on the simple triangle graph.

Let us take an equilateral triangle, and associate to its vertices the two dimensional complex Hilbert spaces ${\cal H}_S$ , ${\cal H}_T$ and ${\cal H}_U$ of Sarah, Tom and Ursula.
Let us now chose a particular two-qubit state from each of these spaces.
First let us define 

\beq
{\vert \pm\rangle}_S=\frac{1}{\sqrt{2}}({\vert 0\rangle}_S\pm{\vert 1\rangle}_S),   \quad
{\vert \pm\rangle}_T=\frac{1}{\sqrt{2}}({\vert 0\rangle}_T\pm{\vert 1\rangle}_T),   \quad                                                                            {\vert \pm\rangle}_U=\frac{1}{\sqrt{2}}({\vert 0\rangle}_U\pm{\vert 1\rangle}_U),
\eeq
\noindent
which are just the Hadamard transformed states $\vert\overline{0}\rangle$ and $\vert\overline{1}\rangle$ of the ones $\vert 0\rangle$ and $\vert 1\rangle$,
and associate to the triangle graph the three-qubit state

\beq
\vert ---\rangle\equiv {\vert -\rangle}_U\otimes{\vert -\rangle}_T\otimes {\vert -\rangle}_S.
\eeq
\noindent
Our graph state is arising by specifying the interactions between the states of the vertices along the three edges of the triangle.
Consider the interactions (for their physical meaning see the Appendix) of the following form

\beq
V_{TS}=I\otimes I\otimes P_++I\otimes {\cal Z}\otimes P_-,\quad
V_{UT}=I\otimes P_+\otimes I+{\cal Z}\otimes P_-\otimes I.\quad
V_{US}=I\otimes I\otimes P_++{\cal Z}\otimes I\otimes P_-,
\eeq
\noindent
where for the definitions of the $2\times 2$ matrices $P_{\pm}$ and ${\cal Z}$ see the Appendix.
Now it is straightforward to check that the graph state

\beq
{\vert G\rangle}_{---}=V_{TS}V_{UT}V_{US}\vert ---\rangle,
\eeq
\noindent
is up to the factor $\sqrt{8}\omega$ is precisely the state of Eq. (78).
Moreover, had we chosen the state

\beq
\vert -++\rangle={\vert -\rangle}_U\otimes {\vert +\rangle}_T\otimes {\vert +\rangle}_S
\eeq
\noindent
as the starting state attached to the corresponding vertices of the triangle graph we would have obtained the other graph state

\beq
{\vert G\rangle}_{-++}=V_{TS}V_{UT}V_{US}\vert -++\rangle,
\eeq
\noindent
which is up to $-\sqrt{8}\omega$ is just the state of Eq. (81) corresponding to the non-BPS class $(-++)$.
The remaining cases are obtained by permutation of the signs $(-++)$.
Hence we managed to demonstrate that the entangled states corresponding to non-BPS black hole solutions for the $D2-D6$ system characterized by the condition $Z\neq 0$ are just graph states associated to the simple triangle graph.

\subsection{N=8 reinterpretation of the STU-model. Density matrices.}

In this subsection using some more results from quantum information theory we would like to comment on the embedding of the solutions of the  $d=4$, $N=2$ STU-model in $d=4$, $N=8$ supergravity\cite{FK2}.
As we have seen the $Z\neq 0$ extremal black-hole solutions of the STU-model
can be given a nice reinterpretation in terms of a moduli and charge-dependent  {\it pure} three-qubit entangled state.  
How to describe the embedding of these solutions in the ones of $N=8$ supergravity?
One way to do this is to consider the {\it pure} state tripartite entanglement of seven qubits\cite{Ferrara,Levay2,Duff2}.
There is however, another possibility to describe the solutions in the $N=8$ context using {\it mixed} three-qubit states, characterized by a density matrix with special properties.

The main idea is to associate the matrix of the central charge $Z_{AB}, A, B=0,1,\dots 7$ to a {\it bipartite} system consisting of two indistinguishable fermionic subsystems with $2M=N=8$ single-particle states. This system is characterized by the pure state

\beq
\vert\chi\rangle=\sum_{A,B=0}^{2M-1}Z_{AB}{\hat{c}_A}^{\dagger}{\hat{c}_B}^{\dagger}\vert \Omega\rangle\in{\cal A}({\bf C}^{2M}\otimes {\bf C}^{2M})
\eeq
\noindent
where
\beq
\{\hat{c}_A,{\hat{c}_B}^{\dagger}\}={\delta}_{AB},\quad \{\hat{c}_A,\hat{c}_B\}=0,\quad \{{\hat{c}_A}^{\dagger},{\hat{c}_B}^{\dagger}\}=0,\quad A,B=0,\dots 2M-1.
\eeq
\noindent
Here $Z$ is a $2M\times 2M$ complex antisymmetric matrix, $\hat{c}_A$ and ${\hat{c}_A}^{\dagger}$ are fermionic annihilation and creation operators, $ \vert\Omega\rangle$ is the fermionic vacuum and the symbol ${\cal A}$ refers to antisymmetrisation\cite{Schliemann,Levay3}.
It can be shown\cite{Schliemann} that the normalization condition $\langle\chi\vert\chi\rangle =1$ implies that $2{\rm Tr}ZZ^{\dagger} =1$.
However, since our states in the black hole analogy are unnormalized we do not need this condition.

As was demonstrated in the literature\cite{Schliemann} local unitary transformations $U\otimes U$ with $U\in U(2M)$ acting on ${\bf C}^{2M}\otimes {\bf C}^{2M}$ do not change the fermionic correlations and under such transformations $Z$ transforms as

\beq
Z\mapsto UZU^T.
\eeq
\noindent
In the black hole context for $2M=N=8$ the group $U(8)$ is the automorphism group of the $N=8$ , $d=4$ supersymmetry algebra.

Since the fermions are indistinguishable, the reduced one-particle density matrices are equal and have the form\cite{Paskauskas}

\beq
{\rho}=ZZ^{\dagger}.
\eeq
\noindent
(For normalized states we have $\rho=2ZZ^{\dagger}$ in order to have ${\rm Tr}\rho =1$. However, for unnormalized states, our concern here, we prefer to swallow the factor of $2$ in Eq. (94).)
However now we cannot pretend that any of the one-particle density matrices describes the properties of precisely the first or the second subsystem.
$\rho$ describes the properties of a randomly chosen subsystem that cannot be better identified\cite{Ghirardi}.
A useful measure describing fermionic entanglement for $M=2$ (which correspond to N=4 supergravity) is\cite{Schliemann,Levay3}

\beq
\eta\equiv 8\vert Z_{01}Z_{23}-Z_{02}Z_{13}+Z_{03}Z_{12}\vert=8\vert {\rm Pf}(Z)\vert.
\eeq
\noindent
For normalized states $0\leq \eta\leq 1$. For $M>2$ similar measures related to the Pfaffian in higher dimensions have also been considered in the literature\cite{Sch2}.
A fermionic analogue of the usual Schmidt decomposition can also be introduced.
According to this result\cite{Schliemann} (which is just a reinterpretation of an old result of Zumino\cite{Zumino}) there exists an unitary matrix ${\cal U}\in U(2M)$
such that

\beq
{\Lambda}={\cal U}Z{\cal U}^T, \qquad {\Lambda}=\bigoplus_{j=0}^{M-1} {\zeta}_j{\varepsilon},\quad \varepsilon=\begin{pmatrix}0&1\\-1&0\end{pmatrix}.
\eeq
\noindent
The number of {\it nonzero} complex numbers ${\zeta}_j, j=0,1,\dots M-1$ is called the {\it Slater rank} of the fermionic state. A fermionic state is called {\it entangled} if its Slater rank is greater than $1$.
For $M=2$ a sufficient and necessary condition for having Slater rank $1$ states is the vanishing of $\eta$ i.e. the Pfaffian of $Z$ (see Eq. (95)). For $M>2$ similar conditions can be found in Refs. 39. and 41.
Such states can always be written in terms of one Slater determinant, i.e. in this case $Z_{AB}$ is a {\it separable bivector}.   
Note, that to the process of obtaining the block diagonal form Eq. (96) in the black hole picture corresponds the one of finding the canonical form of the central charge matrix $Z_{AB}$.

One can alternatively characterize bipartite entanglement by the entropies of von Neumann and of R\'enyi\cite{Levay3}

\beq
S_1=-{\rm Tr}\rho{\log}_2\rho,\qquad S_{\alpha}=\frac{1}{1-\alpha}{\log}_2{\rm Tr}{\rho}^{\alpha},\quad \alpha>1.
\eeq
\noindent
Here the von Neumann entropy $S_1$ is the $\alpha\to 1$ limit of Renyi's ${\alpha}$ entropies.
For femionic states one calculates the eigenvalues $\vert{\zeta}_j\vert^2$ of the reduced density matrix $\rho$ of Eq. (94).
Then the entropies have the form

\beq
S_1=1-\sum_{j=0}^{M-1}\vert{\zeta}_j\vert^2{\log}_2\vert{\zeta}_j\vert^2,\quad
S_{\alpha}=1+\frac{1}{1-\alpha}\sum_{j=0}^{M-1}\vert{\zeta}_j\vert^{2\alpha}.
\eeq
\noindent
The fact that for fermionic systems these entropies satisfy the bound $1\leq S_{\alpha}$ can be traced back to the fact that for fermionic density matrices the so-called generalized Pauli principle holds\cite{Levay3}.
This is to be contrasted with the bound $0\leq S_{\alpha}$ which holds for bipartite systems with distinguishable subsystems.
Some special cases of $S_{\alpha}$ are often encountered, for exmaple the quantity

\beq
{\rm Tr}\rho^2=\sum_{j=0}^{M-1}\vert{\zeta}_j\vert^4,
\eeq
\noindent
is called {\it the purity} of the mixed state $\rho$. Obviously one has
$S_2=-{\log}_2[{\rm Tr}\rho^2]$. 

Let us now consider the central charge matrix in the $N=8$ theory

\beq
Z_{AB}=f_{AB}^{\Lambda\Sigma}Q_{\Lambda\Sigma}-h_{\Lambda\Sigma,AB}P^{\Lambda\Sigma},\qquad A,B,\Lambda,\Sigma=1,2,\dots 8,\quad A<B,\quad \Lambda<\Sigma
\eeq
\noindent
where the charge vector $(Q^{\Lambda\Sigma},P_{\Lambda\Sigma})$
is forming the fundamental representation of $E_{7(7)}$.
The beins $f_{AB}^{\Lambda\Sigma}(\phi)$ and $h_{AB,\Lambda\Sigma}(\phi)$ are depending on the $70$ scalar fields of the coset $E_{7(7)}/SU(8)$.
The black hole potential for $N=8$, $d=4$ supergravity has the following form\cite{FK2}

\beq
{\cal V}_{BH}(\phi;Q,P)=Z_{AB}\overline{Z}^{AB}={\rm Tr}ZZ^{\dagger}=\frac{1}{2}{\rm Tr} \rho,
\eeq
\noindent
where subscripts $A,B$ label an ${\bf 8}$ and superscripts label an $\overline{\bf 8}$ of $SU(8)$. Hence ${\overline{Z}}^{AB}$ refers to the complex conjugate of the central charge. (Summation is understood only for $A<B$.)
Notice that using Eq. (94) we have also introduced the (unnormalized) reduced density matrix. It is hermitian $\rho ={\rho}^{\dagger}$, positive $\rho\geq 0$, however now it is {\it not} satisfying the additional normalization condition ${\rm Tr}\rho =1$. 
Eq. (101) has to be compared with our previous result of Eq. (30).
Both of these equations express the black hole potential as half of the "norm" of a moduli and charge dependent state. 
However, for the $N=8$ case it is a {\it mixed state}.
Since the $N=2$, STU model can be regarded as a consistent truncation of the $N=8$ case, one might suspect that the mixed state $\rho$ is somehow related to the pure one $\Psi$ of Eq. (26). Using the result of Ref. 20 we can easily establish the desired relationship.
Indeed it has been shown that the algebraic attractor equations of the $N=8$ theory can be identified with the corresponding $N=2$ attractor equations, under
the correspondence

\beq
\zeta_0=iZ,\qquad \zeta_1=\overline{D_{\hat{S}}Z},
\qquad\zeta_2=\overline{D_{\hat{T}}Z},\qquad\zeta_3=\overline{D_{\hat{U}}Z}.
\eeq
\noindent
Using Eqs. (16), (31), (36) and (42) we can identify these with the components of $\vert{\Psi}\rangle$ of Eq. (26) as

\beq
i\zeta_0={\Psi}^{111},\qquad i\zeta_1={\Psi}^{001},\qquad i\zeta_2={\Psi}^{010},\qquad i\zeta_3={\Psi}^{100}.
\eeq
\noindent
Notice also that these components are related to the remaining ones by Eq. (36)
. 

Now we use instead of the labeling $A,B=0,1,\dots 7$ the binary one of $000,001,\dots ,111$ to write the density matrix in the form 

\beq
\rho=
\sum_{lkj=0}^1{\vert{\Psi}^{lkj}\vert}^2\vert lkj\rangle\langle lkj\vert=\vert Z\vert^2[\vert 000\rangle\langle 000\vert +\vert 111\rangle\langle 111\vert]+\dots
\eeq
\noindent
Here the vectors $\vert lkj\rangle$ are the {\it eigenvectors} of the matrix $ZZ^{\dagger}$ depending on the remaining charges and moduli.
In this way we managed to represent $\rho$ as a mixed state, where the $8$ weights appearing in the mixture are determined by the $8$ moduli-dependent amplitudes of the pure state of the  $STU$ model.
They are multiplying the three-qubit pure states $\vert lkj\rangle\langle lkj \vert$ the mixture is composed of.
However, a density matrix can be written in many different ways as the
convex linear combination of different types of pure states\cite{Hugh}.
The one based on the eigenvectors of $\rho$ is just one of them.
Of course the "quantum" ensembles to be considered here has to be chosen from
the subclass compatible with the $U$-duality group $E_{7(7)}$.  
It would be nice to establish an explicit correspondence between consistent truncations\cite{Ferrara,Levay2} of the $N=8$ model other than the $N=2$ STU one and these alternative
decompositions of $\rho$.

The possibility of interpreting $\rho=ZZ^{\dagger}$ as a mixed three-qubit state depending on the $56$ charges and $70$ moduli fields has further illuminating aspects.
It is well-known that the entropy formula for regular $N=8$ black holes in four dimensions can be given in terms of the square root of the magnitude of the unique Cartan-Cremmer-Julia quartic invariant $J_4$\cite{Julia2,Cartan,FK2} constructed from the fundamental ${\bf 56}$ of the group $E_{7(7)}$.
Using the definition of $\rho$ $J_4$ can be expressed as

\beq
J_4={\rm Tr}{\rho^2}-\frac{1}{4}({\rm Tr}\rho)^2+8{\rm Re}{\rm Pf}(Z), \qquad \rho=ZZ^{\dagger}.
\eeq
\noindent
Notice that the terms contributing to $J_4$ are the {\it purity} Eq. (99) (which is related to Renyi's entropy $S_2$), one fourth of the the norm squared and $8$ times the real part of a quantity similar to the fermionic entanglement measure $\eta$ of Eq. (95).
All these terms are invariant under the group of local unitary transformations which is $U(8)$. However, their particular combination is invariant under the larger group $E_{7(7)}$ as well. 
It is tempting to interpret $J_4$ as an entanglement measure for a {\it special}
subclass of three-qubit mixed states. Apart from the fact that $\rho$ is an $8\times 8$ matrix the three-qubit reinterpretation is also justified by the recent
reinterpretation of the ${\bf 56}$ of $E_7$ in terms of seven three-qubit states\cite{Ferrara,Levay2,Duff2}. We note in this context that finding a suitable measure of entanglement for mixed states is a difficult problem. We remark that the only explicit formula known is the celebrated one of Wootters for two-qubit mixed states\cite{Wootters}.
$J_4$ might possibly serve as an entanglement measure 
for three-qubit mixed states having doubly degenerate eigenvalues which is related to the fact that the purification of $\rho$ is the fermionic entangled state
of Eq. (91).

Using this density matrix picture let us now look at the BPS and non-BPS solutions as embedded in the corresponding $N=8$ ones.
According to Eqs. (62), (102-103) for the BPS case
we have

\beq
\zeta_0\neq 0,\quad \zeta_1=\zeta_2=\zeta_3=0,\qquad {\cal S}=\pi\vert Z\vert^2_{BPS},
\eeq
\noindent
where the central charge is calculated at the attractor point.
The corresponding density matrix has the form

\beq
\rho_{BPS}=\vert Z\vert^2_{BPS}\{\vert 000\rangle\langle 000\vert+\vert 111\rangle\langle 111\vert\},
\eeq
\noindent
which is a state of {\it Slater rank 1}. Hence for BPS states the corresponding fermionic purification Eq. (91) can be expressed using $Z_{AB}$ as a separable bivector. The arising state of Eq. (91) is a one consisting of merely one Slater determinant expressed in terms of two states with eight single particle states.

For the non-BPS case we have

\beq
\vert\zeta_0\vert=\vert\zeta_1\vert=\vert\zeta_2\vert=\vert\zeta_3\vert\neq 0,\qquad {\cal S}=4\pi\vert Z\vert^2_{nonBPS}
\eeq
\noindent
with the corresponding mixed state

\beq
\rho=\vert Z\vert^2\sum_{lkj}\vert lkj\rangle\langle lkj\vert
\eeq
\noindent
which is a state of Slater rank $4$.
According to Schliemann et.al.\cite{Schliemann} a fermionic state is called entangled if and only if its Slater rank is strictly greater than $1$.
Hence class I. solutions correspond to non-entangled, and class II. solutions
correspont to entangled fermionic purifications.
We have to be careful however, not to conclude that BPS solutions are represented by non-entangled fermionic purifications and non-BPS solutions with entangled ones. 
This is because we have not analysed solutions of class III. namely the ones with $Z\equiv 0$. For these solutions we have\cite{Bellucci3}
for example

\beq
\zeta_0=0,\qquad \zeta_1\neq 0,\qquad \zeta_2=0,\qquad \zeta_3=0,
\eeq
\noindent
hence these solutions also give rise to Slater rank one (i.e. non-entangled ) states .
This is because from the $N=8$ perspective $N=2$ non-BPS $Z=0$ solutions
are originated from the $N=2$ BPS ones by simply exchanging the eigenvalues
$\vert\zeta_0\vert^2$  and $\vert\zeta_1\vert^2$ of $\rho$.

\section{Discussion}

In order to make the picture complete, let us also comment on the non-BPS solutions of type III. for the $D2-D6$ system.
For such solutions we have $Z\equiv 0$. Let us chose the signs for the charges as follows\cite{Ceresole2} 

\beq
p^0<0,\qquad q_3<0,\qquad q_2<0,\qquad q_1>0.
\eeq
\noindent
For this combination the solutions are\cite{Bellucci3}

\beq
S=-\frac{i}{q_1}\lambda,\qquad T=\frac{i}{q_2}\lambda,\qquad U=\frac{i}{q_3}\lambda,\qquad \lambda=\sqrt{-\frac{q_1q_2q_3}{p^0}}.
\eeq
\noindent
A calculation of the three-qubit entangled state Eq. (26) using Eqs. (31-34)
shows that

\beq
{\vert{\Psi}\rangle}_{q_1>0} =i\sqrt{2}(-p^0q_1q_2q_3)^{1/4}(\vert 001\rangle +\vert 110\rangle).
\eeq
\noindent
It is a GHZ-like state obtained from the canonical GHZ state corresponding to the BPS solutions (see Eq. (68)) by  applying the bit flip error operation
$I\otimes I\otimes X$.
This is consistent with our interpretation that non-vanishing covariant derivatives of $Z$ at the attractor point (in this case  $D_SZ\neq 0$) are represented by bit flip errors.
By permutation symmetry the remaining two cases with the sign of $q_2$ and then the sign of $q_3$ is chosen to be positive will result in the states

\beq
{\vert{\Psi}\rangle}_{q_2>0} =i\sqrt{2}(-p^0q_1q_2q_3)^{1/4}(\vert 010\rangle +\vert 101\rangle),
\eeq
\noindent
and
\beq
{\vert{\Psi}\rangle}_{q_3>0} =i\sqrt{2}(-p^0q_1q_2q_3)^{1/4}(\vert 100\rangle +\vert 011\rangle),
\eeq
\noindent
corresponding to bit flip errors $I\otimes X\otimes I$ and $X\otimes I\otimes I$ ($D_TZ\neq 0$ and $D_UZ\neq 0$).

In the Hadamard transformed basis the connection between sign flip errors of charges and phase flip errors is displayed  explicitly.
In this case we have

\beq
{\vert{\Psi}\rangle}_{q_1>0}=i(-p^0q_1q_2q_3)^{1/4}\{\vert\overline{000}\rangle-\vert\overline{011}\rangle-\vert\overline{101}\rangle+\vert\overline{110}\rangle\}.
\eeq
\noindent
Comparing this with the corresponding state for the BPS solution 
($p^0<0, q_3>0, q_2>0, q_1>0$)

\beq
{\vert{\Psi}\rangle}_{q_3>0,q_2>0,q_1>0}=i(-p^0q_1q_2q_3)^{1/4}\{\vert\overline{000}\rangle+\vert\overline{011}\rangle+\vert\overline{101}\rangle+\vert\overline{110}\rangle\}.
\eeq
\noindent
and the reference state
\beq
\vert\psi\rangle=p^0\vert 000\rangle +q_3\vert 011\rangle +q_2\vert 101\rangle +q_1\vert 110\rangle,
\eeq
\noindent
clearly shows that at the attractor point the phase flip error $I\otimes I\otimes {\cal Z}$ in the Hadamard transformed base transforming the BPS solution to the non-BPS $Z=0$ one corresponds to a simultaneous sign flip in the charges $q_2$ and $q_3$.

Now we realize that there is a possibility to present a unified formalism
for the characterization of all extremal black hole solutions found for the
$D2-D6$ system. 
In order to do this let us call the charge configuration related to the BPS case the {\it standard} one.
Hence for the standard configuration we have

\beq
p^0<0,\qquad q_3>0,\qquad q_2>0,\qquad q_1>0.
\eeq
\noindent
Our aim is to describe all the remaining classes of solutions as deviations from this one. This viewpoint is justified by the fact that for the BPS solutions bit flip errors corresponding to sign changes of charges are supressed, but for the remaining non-BPS cases they are not.
Let us define a map

\beq
({\rm sgn}(p^0),{\rm sgn}(q_3), {\rm sgn}(q_2), {\rm sgn}(q_1))\mapsto (d,c,b,a),\qquad d,c,b,a=0,1
\eeq
\noindent
in the following way. For the standard configuration we define
$(dcba)\equiv (0000)$. The occurrence of $1$s in some of the slots is indicating a sign flip of the correspondig charge with respect to the standard configuration. Hence for example the class label $(0110)$                                  refers to the state of Eq. (116) with the signs of the charges $q_3$ and $q_2$
have been changed. 
(Compare Eqs. (116) and (117).)
Then recalling our result in Eq. (83)
for the remaining non-BPS classes of solutions we define 

\beq
{\vert{\Psi}(0)\rangle}_{dcba}=i[-(-1)^dp^0q_1q_2q_3]^{1/4}
\{e^{id\pi}\vert\overline{000}\rangle+e^{ic\pi}\vert \overline{011}\rangle
+e^{ib\pi}\vert \overline{101}\rangle + e^{ia\pi}\vert\overline{110}\rangle\}.
\eeq
\noindent
BPS (class I.) solutions have the label $(0000)$, no charge flips.
For class II. solutions a quick check shows that the class $(1000)$ corresponds to our state of Eq. (80), and the one with label $(1011)$ to the one of Eq. (82). This class can be characterized with and {\it odd} number of sign flips.
In the first case only one charge has been flipped ($p^0$), in the second three ($p^0$, $q_2$ and $q_1$).  
For class III. (non-BPS, $Z=0$) solutions correspond to states like Eq. (116)
with class label $(0110)$. They have an even number of sign flips.

Notice also that classes I. and III. have the same charge orbit structure\cite{Bellucci2} (that correspond to two separated branches of a disconnected manifold)and both of them have an even number of charge flip errors.
Class II. solutions have two subclasses. Its is also useful to recall that the configuration with one charge error is known to be upliftable to a $d=5$ BPS solution, and the other ones with three errors
are upliftable to $d=5$ non-BPS ones\cite{Ceresole2}. 

There is however an important distinction to be made between charge flip errors and bit flip errors. We have a nice correspondence between bit flip errors and sign flip ones only for the $D2$-brane charges $q_1,q_2$ and $q_3$.
The sign flip error of the $D6$-brane charge $p^0$ cannot be understand
in terms of quantum information theory
within the STU model. However, by embedding this model into the $N=8$ one we have seen that the class $p^0<0$ corresponds to mixed states with fermionic purification having Slater rank {\it one}, and $p^0>0$ with fermionic purifications having Slater rank {\it four}.  
Notice also that from Eqs. (30) and (121) we immediately obtain the result (see Eq. (4.20) of Ref. 19.)

\beq
V_{BH}(0)=2{\vert p^0q_1q_2q_3\vert}^{1/2}=\sqrt{\vert D\vert}.
\eeq
\noindent

We remark that the dual situation for extremal black hole solutions (i.e. the $D0-D4$) system is showing similar features. In this case states very similar to the ones of
Eq. (121) can be introduced. This classs of states will contain the basis
states $\vert\overline{111}\rangle$, $\vert\overline{100}\rangle$, $\vert\overline{010}\rangle$ and $\vert\overline{001}\rangle$, i.e. states with opposite parity than the ones of Eq. (121).
Of course our interpretation in terms of charge and bit flip errors still survives.

Let us summarize the main results found in this paper.
For the STU model we have introduced a three-qubit entangled {\it pure} state which is depending on the charges and the moduli. The different components of this pure state are obtained by replacing in the superpotential $W$ some of the moduli with their complex conjugates (see Eqs. (31-34)). 
In terms of this unnormalized pure state the black hole potential can be written as one half the norm of this state (Eq. (30)).
The flat covariant derivatives with respect to the moduli are acting
on this pure state as bit flip errors (Eqs. (45-48)).
In other words: the representatives of the flat moduli dependent covariant derivatives at the attractor point are the bit flip errors.
Using our entangled state BPS and non-BPS ($Z=0$) solutions can be characterized
 as the ones containing only $GHZ$ components or their bit-flipped versions
(Eqs. (63-64), (68) and (113-115)).   
For non-BPS ($Z\neq 0$) solutions the corresponding states have amplitudes with equal magnitudes (Eq. (58)), meaning that these states are linear combinations
of all states with suitable phase factors as expansion coefficients.
For the $D2-D6$ (and its dual $D0-D4$) systems the expansion coefficients are 
just positive or negative signs, and the corresponding states are graph states.
For the $D2-D6$ ($D0-D4$) systems in the Hadamard (discrete Fourier ) transformed base the states at the attractor point show a universal behaviour (see Eq. (121).
The bit flip errors in this base correspond to phase flip ones, which correspond to the sign flips of the charges $q_1,q_2,q_3$, ($p^1,p^2,p^3$).
For BPS solutions bit flips are supressed, but for non-BPS solutions they       are not. 
We managed to embed the $N=2$ STU model characterized by a three-qubit {\it pure} state to the $N=8$ one characterized by a three-qubit {\it mixed} one.
These mixed states have femionic purifications.
Fermionic purifications with Slater rank one describe BPS, and the ones with Slater rank four non-BPS solutions.
In the STU truncation for  the $D2-D6$ system these classes correspond to the 
charge configurations with $p^0<0$ and $p^0>0$ respectively.
Finally we remark that the nice universal behavior we have found (see Eq. (121) and its $D0-D4$ analogue) are also physically relevant cases that correspond to {\it stable} minima of $V_{BH}$\cite{Bellucci2,Kallosh,Triv2}, i.e. they are all attractors in a strict sense.

\section{Acknowledgements} 
Financial support from the Orsz\'agos Tudom\'anyos Kutat\'asi Alap              (grant numbers T047035, T047041, T038191) is                                    gratefully acknowledged.

\section{Appendix}

In this appendix we briefly summarize some background material from quantum information theory needed in the main body of the paper.
The qubit is an element of a two complex dimensional vector space ${\bf C}^2$
with basis vectors (computational base) denoted by $\vert 0\rangle$ and $\vert 1\rangle$. These correspond to the usual basis vectors that are eigenvectors of the Pauli matrix $\sigma_3$. This operator is conventionally denoted by ${\cal Z}$ (not to be confused with the central charge $Z$) and is called the phase flip operator.
Hence we have

\beq
{\cal Z}\vert 0\rangle =\vert 0\rangle,\qquad {\cal Z}\vert 1 \rangle = -\vert 1\rangle.
\eeq
\noindent
The Pauli matrix $\sigma_1$ (conventionally denoted by $X$) is used to represent bit flips

\beq
X\vert 0\rangle =\vert 1\rangle,\qquad X\vert 1\rangle =\vert 0\rangle.
\eeq
\noindent
The orthogonal projectors $P_{\pm}$ are defined as

\beq
P_{\pm}=\frac{1}{2}(I\pm {\cal Z}),
\eeq
\noindent
where $I$ is the $2\times 2$ unit matrix.
Systems with $n$ qubits  are defined to be the elements of the  $n$-fold tensor product ${\bf C}^2\otimes {\bf C}^2\otimes\cdots \otimes {\bf C}^2$.

In quantum information theory, especially in quantum error correction
the discrete Fourier or Hadamard transformed base is often used.
The Hadamard transformed basis vectors are denoted by $\vert\overline{0}\rangle$ and $\vert\overline{1}\rangle$ and defined as 

\beq
\vert\overline{0}\rangle =\frac{1}{\sqrt{2}}(\vert 0\rangle +\vert 1\rangle),\qquad \vert\overline{1}\rangle=\frac{1}{\sqrt{2}}(\vert 0\rangle -\vert 1\rangle).
\eeq
\noindent
They are sometimes alternatively denoted by $\vert +\rangle$ and $\vert -\rangle$ since they are eigenvectors of the bit flip operator $X$ with eigenvalues $\pm 1$.
These basis vectors can also be defined by introducing the unitary operator
of Hadamard transformation

\beq
\vert \overline{0}\rangle=H\vert 0\rangle,\qquad \vert\overline{1}\rangle =H\vert 1\rangle,\qquad i.e.\quad H=\frac{1}{\sqrt{2}}\begin{pmatrix}1&1\\1&-1\end{pmatrix}.
\eeq
\noindent
Since $HXH={\cal Z}$ and $H{\cal Z}H=X$ the operator $X$ is acting on the Hadamard transformed base as a phase flip operator and vice versa.
The important corollary of this observation is that in the theory of quantum error correction once we have found a means for correcting bit flip errors using a discrete Fourier transform the same technique can be used for correcting phase flip ones.

For three-qubit systems we use the Hadamard transformation $H^{\otimes 3}=H\otimes H\otimes H$ represented by the $8\times 8$ matrix

\beq
H^{\otimes 3}=\frac{1}{\sqrt{8}}\begin{pmatrix} 1& 1& 1& 1& 1& 1& 1& 1\\
                                                1&-1& 1&-1& 1&-1& 1&-1\\
 1& 1&-1&-1& 1& 1&-1&-1&\\ 1&-1&-1& 1& 1&-1&-1& 1\\1& 1& 1& 1&-1&-1&-1&-1\\ 1&-1& 1&-1&-1& 1&-1& 1\\ 1& 1&-1&-1&-1&-1& 1& 1&\\ 1&-1&-1& 1&-1& 1& 1&-1\end{pmatrix}. 
\eeq
\noindent
Labelling the rows of this matrix in the binary form $(000,001,010,011,100,101,110,111)$ one can verify that we have for example 

\beq
\vert\overline{110}\rangle =\frac{1}{\sqrt{8}}(\vert 000\rangle+\vert001\rangle-\vert 010\rangle -\vert 011\rangle -\vert 100\rangle -\vert 101\rangle +\vert 110\rangle +\vert 111\rangle)
\eeq
\noindent
coming from the sign combinations of the sixth row.
Adding and substracting the first and last rows of the matrix $H^{\otimes 3}$
reveals that

\beq
\frac{1}{\sqrt{2}}(\vert 000\rangle +\vert 111\rangle) =\frac{1}{2}(\vert \overline{000}\rangle +              \vert \overline{011}\rangle +\vert\overline{101}\rangle +                       \vert\overline{110}\rangle)
\eeq
\noindent

\beq
\frac{1}{\sqrt{2}}(\vert 000\rangle -\vert 111\rangle) =\frac{1}{2}(\vert 
\overline{111}\rangle +              \vert \overline{100}\rangle +\vert\overline{010} \rangle +
\vert\overline{001}\rangle).
\eeq
\noindent
This shows that the {relative phase} of the states $\vert 000\rangle $ and $\vert 111\rangle$ in a multipartite superposition can be detected in the Hadamard transformed base via a parity check (in Eq. (130) the number of $1$'s is {\it even} and in Eq. (131) it is {\it odd}).
This is a crucial observation for developing quantum error correcting codes\cite{Steane1,Steane2}. 
Notice that for the $D2-D6$ system at the black hole horizon we have the first and for the $D0-D4$ system the second type of entangled state.
For the general $D0-D2-D4-D6$ system we have a superposition of ${\it even}$ and ${\it odd}$ states (see Eq. (64) and its Hadamard transform).

Let us now introduce the notion of a {\it graph state}.
Consider a simple graph $G$ which contains neither loops nor multiple edges.
Let us denote its vertices by $V$ its edges by $E$.
The main idea is to prepare $n$-qubits (their number is the same as the number of vertices $V$) in some initial vector $\vert\psi\rangle\in {\bf C}^2\otimes\cdots\otimes {\bf C}^2$ 
and then couple them according to some interaction pattern represented by $G$.
It turns out that the interaction pattern can be completely specified by $G$
if it is of the form

\beq
U^I_{xy}=e^{-ig_{xy}{\cal H}^I_{xy}},\qquad {\cal H}^I_{xy}=I\otimes\cdots I\otimes {\cal Z}_x\otimes I\cdots I\otimes {\cal Z}_y\otimes I\cdots\otimes I.
\eeq
\noindent
Here $g_{xy}$ are coupling constants which are the same for every pair of vertices $x,y\in V$. Notice that ${\cal H}^I_{xy}$ is the Ising Hamiltonian operating only between the vertices $x,y\in V$ that are linked by an edge $E$.
The statement is\cite{Dur} that the interaction pattern assigned to the graph $G$ in which the qubits interact according to some two-particle unitaries chosen from a commuting set of interactions is up to phase factors and local ${\cal Z}$ rotations is the one of Eq. (132).
In the theory of graph states the initial state $\vert\psi\rangle$ associated to the graph $G$ is a separable one which is usually of the form
$\vert\overline{00\cdots 0}\rangle$, but we can choose any one of the $2^n$
Hadamard transformed basis vectors.
For convenience the unitary operators appearing in the definition of a          graph state are not the ones of Eq. (132) but rather the combination  

\beq
U_{xy}(g_{xy})=e^{-ig_{xy}/4}e^{ig_{xy}{\cal Z}_x/4}e^{ig_{xy}{\cal Z}_y/4}U^I_{xy}(g_{xy}/4).
\eeq
\noindent
Here the operators having a particular label are merely acting on the corresponding qubit, on the remaining ones the unit matrix is operating. Notice that $U_{xy}$ is the same as the Ising one up to a phase and two ${\cal Z}$ rotations.
The unitary $U_{xy}$ with $g_{xy}\equiv \pi$ is just the controlled phase gate

\beq
{\cal U}_{xy}\equiv U_{xy}(\pi)=(P_+)_x\otimes I_y+(P_-)_x\otimes {\cal Z}_y=\begin{pmatrix}1&0&0&0\\0&1&0&0\\0&0&1&0\\0&0&0&-1\end{pmatrix},
\eeq
\noindent
where only the relevant $4\times 4$ part of the $2^n\times 2^n$ matrix was displayed.
Now we can define a graph state as

\beq
\vert G\rangle \equiv \prod_{\{x,y\}\in E}{\cal U}_{x,y}\vert\overline{00\cdots 0}\rangle=
\prod_{\{x,y\}\in E}{\cal U}_{x,y}\vert ++\cdots +\rangle,
\eeq
\noindent
Such states for $n=3$  based on the triangle graph are appearing in Eqs. (88)   and (90) in connection with non-BPS solutions with $Z\neq 0$.                    Notice, however that in these cases unlike in the                                   definition of Eq. (135) the qubits are labelled from the right to the left
(compare Eqs. (87) and (134)). Moreover, the states to be entangled with respect to the graph $G$ are the ones of $\vert\overline{111}\rangle=\vert ---\rangle$
and $\vert\overline{100}\rangle =\vert -++\rangle$, rather then the conventional one $\vert\overline{000}\rangle=\vert +++\rangle$.

\end{document}